\documentstyle[rmaaconf]{article}
\begin{document}
\newcommand{\beq}{\begin{equation}}
\newcommand{\eeq}{\end{equation}}
\newcommand{\Mdot}{\dot{M}} 
\newcommand{\hii}{H~{\sc ii}~}
\newcommand{\kms}{\mbox{ km s$^{-1}$}}
\newcommand{\cms}{\mbox{ cm s$^{-1}$}}
\newcommand{\gcm}{\mbox{ g cm$^{-3}$}}
\newcommand{\gcms}{\mbox{ g cm$^2$ s$^{-2}$}}
\newcommand{\about}{\mbox{$\sim$}}
\newcommand{\cc}{\mbox{ cm$^3$}}
\newcommand{\Mo}{\mbox{M$_{\odot}$}}
\newcommand{\Lo}{\mbox{$L_{\odot}$}}
\title{INTERACTIONS OF MASSIVE STARS WITH THEIR PARENTAL CLOUDS}

\author{Jos\'e Franco\altaffilmark{1}, Guillermo 
Garc\'{\i}a-Segura\altaffilmark{1}
\& Tomasz Plewa\altaffilmark{2}}
\altaffiltext{1}{Instituto de Astronom\' \i a--UNAM, Apdo. Postal 70-264, 
04510 M\'exico D. F., M\'exico}
\altaffiltext{2}{Max-Planck-Institut f\"ur Astrophysik, Garching, Germany}
\def   \ie   {{\it i.e.},\ }
\def   \eg   {{\it e.g.},\ }
\def   \etal {{\it et al.}\ }
\begin{abstract}
Here we discuss the interaction of massive stars with their parental molecular
clouds. A summary of the dynamical evolution of HII regions and wind-driven 
bubbles in high-pressure cloud cores is given. Both ultracompact HII regions 
and ultracompact wind-driven bubbles can reach pressure equilibrium with their 
surrounding medium. The structures stall their expansion and become static and, 
as long as the ionization sources and the ambient densities remain about 
constant, the resulting regions are stable and long lived. For cases with 
negative density gradients, and depending on the density distribution, some 
regions never reach the static equilibrium condition. For power-law density
stratifications, $\rho \propto r^{-w}$, the properties of the evolution depend 
on a critical exponent, $w_{crit}$, above which the ionization front cannot be 
slowed down by recombinations or new ionizations, and the cloud becomes fully 
ionized. This critical exponent is $w_{crit}=3/2$ during the expansion phase. 
For $w>3/2 $ the gas expands supersonically into the surrounding ionized 
medium, and there are two regimes separated by $w=3$. For $3/2<w\leq 3$, the 
slow regime, the inner region drives a weak shock moving with almost constant
velocity through the cloud. For $w> 3$, the fast regime, the shock becomes 
strong and accelerates with time. Finally, the evolution of slow winds in
highly pressurized region is described briefly.
\end{abstract}
\section{Introduction}

Young stars display vigorous activity and their energy output stirs and heats
the gas in their vicinity. Low-mass stars provide a small energy rate and 
affect only small volumes, but their collective action can provide partial 
support against the collapse of their parental clouds, and could regulate 
some aspects of the cloud evolution (\eg Norman \& Silk 1980; Franco \& Cox 
1983; Franco 1984; McKee 1989). In contrast, stars with initial masses above 8 
\Mo, massive stars, inject large amounts of radiative and mechanical energy 
from their moment of birth until their final explosion as a supernova. In the 
general, low-density, interstellar medium of a gaseous galaxy, the combined 
effects of supernovae, stellar winds, and HII region expansion destroy 
star-forming clouds, produce the hottest gas phases, create large expanding 
bubbles, and are probably responsible for both stimulating and shutting off the 
star formation process at different scales (\eg Cox \& Smith 1974; Salpeter 
1976; McKee \& Ostriker 1977; Franco \& Shore 1984; Cioffi \& Shull 1991; 
Franco \etal 1994; Silich \etal 1996). Thus, the collection of OB associations 
in gaseous galaxies represents a rich energy source which may be controlling 
the general structure of the interstellar medium, and the star formation rate 
(\eg Mueller \& Arnett 1976; Gerola \& Seiden 1978; Franco \& Shore 1984; 
Dopita 1988; see reviews by Tenorio-Tagle \& Bodenheimer 1988, Franco 
1991,1992, Ferrini 1992, and Shore \& Ferrini 1994).

The strong UV radiation field from massive stars create large photoionized, HII
regions. Young HII regions have large pressures and, when the pressure of the
surrounding medium is low, they expand fast and drive a strong shock wave
ahead of the ionization front. Expanding HII regions ionize and stir the 
parental cloud and, when the ionization front encounters a strong negative 
density gradient, they create fast ``champagne" flows and can also generate 
cometary globules and elephant trunks (\eg Tenorio-Tagle 1982; Yorke 1986; 
Franco \etal 1989,1990; Rodriguez-Gaspar \etal 1995; Garc\'{\i}a-Segura \& Franco 
1996). These flows are generated by the pressure difference between the HII 
region and the ambient medium, and they are responsible for the disruption of 
the cloud environment (\eg Whitworth 1979; Elmegreen 1983; Larson 1992; Franco 
\etal 1994). Individual HII regions are bright objects and they are used as
tracers of the active star formation sites in external galaxies (\eg Osterbrock 
1989).

Stellar winds and supernova explosions, on the other hand, being powerful 
sources of mechanical energy generate overpressured regions which drive shock 
waves into the ambient medium (see reviews by 
Ostriker \& McKee 1988, and Bisnovatyi-Kogan \& Silich 1995). The resulting 
wind-driven bubbles and supernova remnants, either from a single progenitor or
from an entire association, are believed to generate most of the structuring 
observed in a gaseous galactic disk (\eg Reynolds \& Ogden 1978; Cowie \etal 
1979; Heiles 1979; McCray \& Snow 1979). Actually, many observed structures in 
the Milky Way and in external galaxies have been ascribed to this stellar 
energy injection (\eg Heiles 1979, 1984; Brinks \& Bajaja 1986; Deul \& Hartog 
1990; Palous \etal 1990, 1994). These bubbles may create fountains or winds at 
galactic scales (\eg Shapiro \& Field 1976; Chevalier \& Oegerle 1979; Bregman 1980; Cox 1981; Heiles 1990; Houck \& Bregman 1990), and their expanding shocks 
have also been suspected of inducing star formation (\eg Herbst \& Assousa 
1977; Dopita \etal 1985). Thus, stellar activity creates a collection of 
cavities with different sizes, and can be viewed as an important element in
defining the structure and activity of star-forming galaxies. Galaxies, 
however, are open systems and their properties are also defined by the
interactions with neighboring galaxies. Here we describe only the effects of 
the stellar energy injection in high-density, high-pressure, regions.

\section{The Parental Clouds and Massive Stars}

Molecular clouds have complex density and velocity distributions, and are 
composed of a variety of high-density condensations. In our Galaxy, they have
non-thermal turbulent velocities, reaching up to about 10 km s$^{-1}$, and 
fairly strong magnetic fields, of up to tens of mG (\eg Myers \& Goodman 1988). 
The $average$ densities for molecular cloud complexes is between $10^2$ and 
$10^3$ cm$^{-3}$, but the high-density condensations have average densities of 
about $\sim 10^6$ cm$^{-3}$ and may even reach values in excess of $10^8$ 
cm$^{-3}$ (\eg Bergin \etal\ 1996; Akeson \etal\ 1996; see review by Walmsley 
1995). These high density condensations, or cloud cores, are the actual sites 
of star formation, and the initial shape and early evolution of the resulting 
HII regions depend on the corresponding core density distributions and 
pressures. Theoretical studies on the collapse of clouds indicate that rotating 
and magnetized cores evolve into flattened (disk-like) structures (\eg 
Bodenheimer \& Black 1978; Cassen \etal 1985), but nonrotating and nonmagnetic 
cases remain spherically symmetric with power-law density distributions (\eg 
Larson 1974). Isothermal spheres in hydrostatic equilibrium have a density 
distribution $\rho \sim r^{-2}$, and the distribution evolves 
towards $r^{-3/2}$ during the free-fall collapse. The pressures at the centers
of these cores are fairly large, and can reach values of about 5 or 6 orders
of magnitude above the pressures at the cloud boundaries (see Garc\'{\i}a-Segura \&
Franco 1996).

Recent observational studies are revealing the structure of the dense star 
forming regions (see review by Walmsley 1995), and provide the average 
parameters of the core density distributions. Radio observations of cloud 
cores and dark clouds, along with visual extinction studies in nearby 
star forming clouds, indicate internal density distributions ranging from 
$r^{-1}$ to $r^{-3}$ (\eg Arquilla \& Goldsmith 1985; Chernicaro \etal 1985; 
Gregorio Hetem \etal 1988; see also Myers 1985). A reasonable mean value for 
the observationally derived power-law distributions is $\rho \sim r^{-2}$.
The typical sizes of the massive high-density cores are about $r_{c}\sim 
0.1$ pc (see Walmsley 1995), and the observationally derived core masses are 
in the range of 10 to 300 M$_{\odot}$ (\eg Snell \etal\ 1993). The observed
properties of ultracompact HII regions (UCHII), on the other hand, indicate 
that the 
exciting stars (one or several massive stars) are embedded in dense and warm
cores, with densities between 10$^4$-10$^7$ cm$^{-3}$ and temperatures of about 10$^2$ K (\eg Churchwell 1990; Cesaroni \etal 1994; Kurtz \etal 1994; Hofner 
\etal 1996; Hurt \etal 1996). The early stages of HII region evolution, then,
occur inside these dense cloud cores. 

\subsection{HII region expansion at constant densities}

Beginning with Str\"omgren (1939) and Kahn (1954), the expansion and evolution
of HII regions has been studied with analytical and numerical models
(see Yorke 1986; Osterbrock 1989; Franco \etal 1989,1990). For a constant 
photon flux and uniform ambient densities, the evolution has well defined 
formation and expansion phases. During the formation phase, the UV photon
field creates an ionization front that moves through the gas. Its speed is
reduced (by geometrical dilution and recombinations), approaching a value of 
about twice the speed of sound in the ionized gas in, approximately, one 
recombination time. At this moment, which marks the end of the formation phase, 
the HII region reaches the initial Str\"omgren size, and the pressure gradient 
across the ionization front begins to drive the expansion of the ionized gas. 
The expansion is supersonic with respect to the surrounding gas and creates a 
shock wave that accelerates and compresses the ambient shocked medium. The 
ionization front sits behind the shock front during the rest of the evolution,
and most of the shocked gas is accumulated in the interphase between the two 
fronts. If either the ionization or the shock front encounters a strong 
negative density gradient (say, the edge of the cloud) and overruns it, then 
the HII region enters into its champagne phase. Numerical models indicate 
that there also appears a strong instability that can fragment the shocked 
shell, and may explain the existence of cometary globules and the highly 
irregular morphologies of ionized nebulae (see Garc\'{\i}a-Segura \& Franco 1996). 

\subsubsection{The Formation Phase}

Assuming a self-gravitating spherical cloud with a molecular density 
distribution that includes a central core, with radius $r_c$ and constant 
density $n_c$, and an isothermal envelope with a power-law density 
stratification $n_{H_2}(r)= n_c{\left(r/r_c \right)}^{-2}$ for $r\geq r_c$.
The total pressure at the core center is
\beq
P(0)=P_0= \frac{2 \pi G}{3} \rho_c^2 r_c^2 + P(r_c)= \frac{8}{5} P(r_c)
\simeq 2\times 10^{-7} \  n_6^2 r_{0.1}^2 \ \ \ \ {\rm dyn \ cm^{-2}},
\eeq
where $G$ is the gravitational constant, $P(r_c)$ is the pressure at
the core boundary $r=r_c$, $n_6=n_c/10^6$ cm$^{-3}$, and
$r_{0.1}=r_c/0.1$ pc. Note that the corresponding molecular mass is
\beq 
M_c\simeq \left( \frac{\pi P_0}{G}\right)^{1/2} r_c^2 \sim 10^2 \
P_7^{1/2} r_{0.1}^2 \ \ \ \ {\rm M_{\odot}},
\eeq
where $P_7= P_0/10^{-7}$ dyn cm$^{-2}$. A star located at the cloud center and 
producing $F_*$ ionizing photons per unit time creates an spherical HII region. 
The gas in the HII region is fully ionized and the ion density is simply 
twice the molecular density, $n_i=2n_{H_2}$. The initial Str\"omgren radius generated by such a star is
\beq
R_s= \left\lbrack 3\ F_*\over {4\pi (2n_c)^2 \alpha_B} \right\rbrack^{1/3}
    \simeq 2\times 10^{-3} F^{1/3}_{48} n^{-2/3}_6 \alpha^{-1/3}_0 \ {\rm pc},
\eeq
where $\alpha_B$ is the hydrogen recombination coefficient to all levels
above the ground level, $\alpha_0=\alpha_B/2.6\times10^{-13}{\rm cm}^3\ {\rm 
s}^{-1}$, $F_{48}=F_*/10^{48}\ {\rm s}^{-1}$, and $n_6=n_c/10^6\ 
{\rm cm}^{-3}$. Thus, $R_s < r_c$ and the initial HII region is well contained 
within the core, and the formation phase follows the well known constant 
density evolution described above. Note that for a dusty cloud, the initial
radius is even smaller and is given by the transcendental equation $R_{S,\rm 
dust}\simeq R_S \ e^{- \tau /3}$ (Franco \etal 1990), where the optical depth 
due to dust absorption is $\tau =\int_0^{R_S, \rm dust} \sigma_{\rm dust} n_0 
dr$, and $\sigma_{\rm dust}$ is the average dust absorption cross-section per 
gas particle. This approximation agrees to better than 10\% with detailed 
radiative transfer calculations (D\'{\i}az \etal 1996). 

The formation phase is completed in a recombination time, and the pressure in 
the ionized region drives a shock into the molecular ambient medium. The HII 
region now begins its expansion phase. The equilibrium temperature in the 
photoionized region, $T_i\sim 10^4$, is achieved in a relatively short time 
scale. The advance of the ionization front is controlled by recombinations in 
the photoionized gas, and the expansion proceeds in a nearly isothermal 
fashion. Neglecting the external pressure, the main features of the expansion 
in a constant density medium can be derived with the thin shell approximation
(see reviews Ostriker \& McKee 1988, Bisnovatyi-Kogan \& Silich 1995, and 
Garc\'{\i}a-Segura \& Franco 1996). Also, the pressureless expansion in density 
stratifications can be solved with simple approximations to the shock front
conditions (Franco \etal 1989,1990). The evolution including the external 
pressure, however, cannot be solved in a closed analytical form, but one can 
derive limits to the main expected features. Obviously, detailed numerical 
simulations provide an adequate description of the evolution in this and more
complicated cases.

\subsubsection{The Expansion Phase}

The evolution can be easily derived by assuming the existence of a 
thin shell with mass $M$, containing all the swept-up ambient gas. This 
approximation can be applied to HII regions when 
the fraction of mass eroded by the ionization front from the shell is small. 
This is true for constant, increasing, or mildly decreasing density 
stratifications, but is not applicable for strongly decreasing gradients 
because the shell is easily eroded by photoionization (Franco \etal 1990). 
Here we consider the constant density case and the thin shell approximation 
can be used without restrictions. Neglecting magnetic fields and self-gravity, 
the equation of motion of the shell, located at a distant $R$ 
from the central star, is
\beq 
4 \pi R^2 (P_i - P_0) = \frac{d}{dt}(M v), \label{newton} 
\eeq
where $P_i$ is the internal pressure, $P_0$ is the external pressure, and $v$ 
is the shell velocity. Assuming that the shell radius can be written as a 
power-law in time, $R=R_0\zeta^{\beta}$ (where $R_0$ is the initial radius of 
the region, $\zeta = (t_0 + t)/t_0$, and $t_0$ is a reference initial time),
with constant $R_0$ or $\beta$ (\ie neglecting the existence of terms with 
$\dot{R_0}$ and $\dot{\beta}$), the right hand side of the equation becomes
\beq
\frac{d}{dt}(M v) = \left(\frac{4 \beta - 1}{3 \beta} \right) 4 \pi
\rho_0 R^2 v^2. 
\eeq
The assumption of a constant $\beta$, as we will see below, does not hold for
cases with $P_0>0$, but the present approximations can still be applied in a 
piece-wise fashion (\ie one can use them in segments, changing the values
for $R_0$ and $t_0$).

The equation of motion can then be written as 
\beq 
P_i - P_0 = \rho_0 v^2 \frac{(4\beta -1)}{3\beta }, 
\eeq
and the shell evolution is given by
\beq 
\frac{dR}{dt}= \left[\frac{3\beta}{(4 \beta -1)}\frac{(P_i - P_0)}{\rho_0} \right]^{1/2}. 
\eeq
Thus, one can write the formal solution to the equation of motion simply as
\beq 
R - R_0 = \int_1^{\zeta} \left[ \frac{3 \beta}{(4 \beta -1) \rho_0} 
\left(P_i-P_0\right) \right]^{1/2} \,\, t_0 \ d\zeta .
\eeq
This formal solution can be applied to HII regions, wind-driven bubbles,
and SN remnants (see below). Obviously, the integration is not straightforward 
unless $P_0=0$, or the pressure difference, $P_i-P_0$, can be written as an explicit function of either $R$ or $t$. In general this is not possible, but 
one can {\it always} check the behavior at early and late times, when the 
external pressure can be neglected and when the internal and external pressures 
become comparable. Here we illustrate the behavior in both cases. 
\subsubsection{Pressure Equilibrium}

\begin{figure}
\vspace*{60mm}
\begin{minipage}{60mm}
\includegraphics{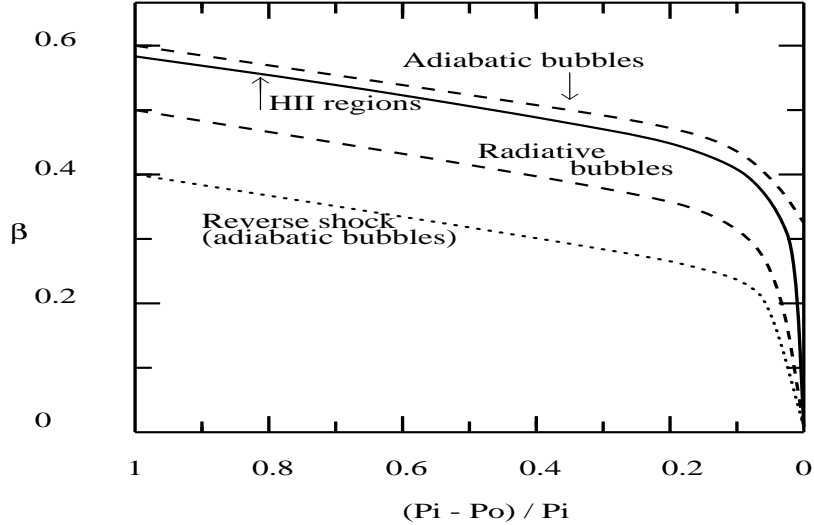}
\end{minipage}

\caption{The evolution of the exponent $\beta$ as a function of the
normalized pressure difference $(P_i -P_0)/P_i$.}
\label{fig:1}
\end{figure}

For the case of expanding HII regions, using $R_0=R_S$ as the initial 
Str\"omgren radius, we can set the solution for the expansion  simply as
\beq 
 R_{HII}(t)=R_S \zeta^{\beta}. 
\eeq
The average ion density inside the HII region at any given time is given by
\beq
<n_i>= \left[ \frac{3\, F_{\star}}{4 \,\pi \,\alpha_B}\right]^{1/2}
R_{HII}^{-3/2},
\eeq
and the corresponding internal pressure is 
\beq
P_i=\left( \frac{3\ k^2 T_i^2 F_{\star}}{\pi\,\alpha_B}
\right)^{1/2} R_{HII}^{-3/2}.
\eeq
The region, then, evolves as
\beq 
R_{HII}= R_S + \int_1^{\zeta} \left( \frac{3 \beta}{(4 \beta -1) \rho_0}
 \left[ \left( \frac{3\,\,k^2 T_i^2F_{\star}}{\pi \alpha_B R_S^3 \zeta^{3\beta}}
\right)^{1/2}- P_0 \right] \right)^{1/2} \,\, t_0 \ d\zeta .
\eeq
At early times, when $P_0/P_i <<1$, the integration with constant $\beta$ gives 
$R_{HII} \ \propto \ \zeta^{1-3\beta /4}$. Thus, our initial assumption for
the power-law implies that $\beta ={1-3\beta /4}$, and one gets $\beta =4/7$.
The solution, then, is now written as
\beq
R_{HII}=R_S \left[ 1 + \frac{7}{4}\left(\frac{8\ k \ T_i}{3\mu_H}\right)^{1/2}
 \frac{t_0}{R_S} (\zeta^{4/7} -1) \right].
\eeq
Defining $c_i\simeq (8\ k \ T_i/3\mu_H)^{1/2}$, and recalling the initial
definition of the power-law $\zeta^{4/7}=R_{HII}/R_S$, the reference time 
is simply given by $t_0=(4/7)(R_S/c_i)$. Thus, as expected, one recovers the 
well known law for HII region expansion in a pressureless medium with a
constant density 
\beq
R_{HII}\simeq R_S \left(1+ \frac{7}{4} \frac{c_i t}{R_S}\right)^{4/7} \ .
\eeq
Using this explicit time dependence, the internal pressure decreases as 
\beq
P_i=P_{i,0} \left(1+ \frac{7}{4} \frac{c_i t}{R_S}\right)^{-6/7},
\eeq
where $P_{i,0}$ is the pressure at $t=0$.

The expansion continues until $P_i\rightarrow P_0$, and the internal pressure 
tends to a constant value. In this limit, the time dependence in the formal 
solution vanishes, giving $3\beta /2 \rightarrow 0$. These limits show that, 
for a constant density medium, $\beta$ evolves as a function of the pressure 
difference from 4/7 to zero. Figure 1 shows the evolution of $\beta$, as a
function of the normalized pressure difference $(P_i-P_0)/P_i$, for HII regions
and wind-driven bubbles (Garc\'{\i}a-Segura \& Franco 1996). The exponents were
derived from high-resolution numerical simulations performed in one dimension 
for the evolution inside high density cores. Figure 2 shows the evolution
of an HII region in a high-density core with $P_7=1$.

When pressure equilibrium is reached, the ion density is simply given by
\beq 
n_{i,{\rm eq}} = \left({P_0 \over 2kT_i}\right)\simeq 3.6 \times 10^4 P_7
T^{-1}_{{\rm HII},4} \ \  {\rm cm^{-3}} \label{ionden},
\eeq
where $P_7 = P_0 / 10^{-7}$ dyn cm$^{-2}$, and $T_{{\rm HII},4} = 
T_i / 10^4$ K. 
The equilibrium radius of the \hii region, then, corresponds to a Str\"omgren
radius at this equilibrium density
\beq 
R_{{\rm S,eq}} \approx 2.9 \times 10^{-2} \,\, F_{48}^{1/3}
\,\, T_{{\rm HII},4}^{2/3} \,\, P_7^{-2/3} \ \  \,\,{\rm pc} , 
\label{Rsequnits} 
\eeq
where $F_{48} = F_{\star}/10^{48}$ s$^{-1}$. 

For high-pressure cores with $r_c\sim 0.1$ pc, the photoionized regions can 
reach pressure equilibrium without breaking out of the core. The resulting
sizes are similar to those of the ultracompact class (see Dyson \etal in
this volume), and indicate that UCHII can be explained by simple pressure
equilibrium. This is in agreement with the recent results reported by Xie \etal 
(1996),
that show the smaller UCHII are embedded in the higher pressure cores. Also
note that the equilibrium values with $P_7 = 1$, $T_{{\rm HII},4}\sim 1$, and
$F_{48}\sim 1$, are {\it very similar} to the average sizes and electron 
densities in UCHII (see Figure 151 of Kurtz \etal 1994). The apparent longevity
problem of UCHIIs is rooted in the notion that young HII regions should grow
fast and reach the expanded state on a relatively short time-scale. This 
statement is false, however, if the external pressure is large and halts the
expansion at a small radius: {\it in pressure equilibrium, UCHIIs are stable 
and long lived}.
\begin{figure}
\vspace*{66mm}
\begin{minipage}{66mm}
\includegraphics{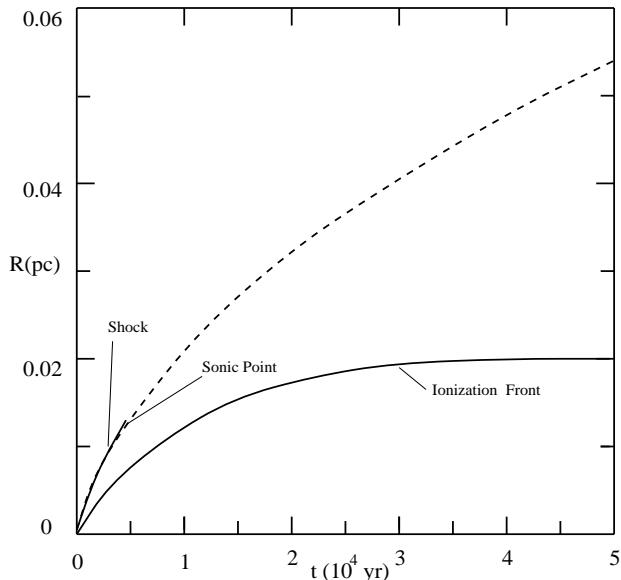}
\end{minipage}

\caption{The evolution of an HII region, with $F_{48} = 1$, in a high-pressure 
core, with $P_7=1$.}
\label{fig:2}
\end{figure}

\subsection{HII Evolution in Decreasing Density Gradients}

If a star is born at a distance smaller than $R_{S, {\rm eq}}$ from the core 
boundary, the speeds of the ionization and shock fronts are modified by the 
negative density gradient. The gas of the HII regions is accelerated in
supersonic flows and no static solution, in pressure equilibrium exists (\eg 
Tenorio-Tagle 1982; Franco \etal 1989, 1990). Thus, under these conditions the 
ultracompact stage is indeed a transient phase. Here we assume isothermal
clouds with $\rho \propto r^{-2}$, but one can easily find the solutions for
the general power-law case, $\rho \propto r^{-w}$ (see Franco \etal 1990). For
$R_s \geq r_c$, the initial ionization front reaches the core radius with a 
speed 
\beq 
 U_c\simeq 90\ \alpha_0 \ n_3 \ r_{17} \left\lbrack
        \left(R_s\over r_c \right)^3-1\right\rbrack
        {\rm km\ s}^{-1}, 
\eeq
and in a time scale
\beq 
 t_c\simeq 130\ \alpha^{-1}_0 n^{-1}_3\ {\rm ln}
        \left\lbrack 1\over 1-(r_c/R_s)^3 \right\rbrack
        \ {\rm yr}, 
\eeq
where $r_{17}=r_c/10^{17}$ cm. Afterwards, the ionization
front enters the density gradient and its speed becomes
\beq 
 U_{if}={U_c \over \left(R_s/ r_c \right)^3-1}u(w),
\eeq
with
\beq 
 u(w)= \cases {\left(r_c/ r_i \right)^{2-w}
         \left\lbrack \left(R_s/ r_c\right)^3 +
         {2w\beta} -{3\beta}\left( r_i/ r_c\right)^{1/\beta}
         \right\rbrack &for $w\not= 3/2$,\cr
         \left(r_c/ r_i \right)^{1/2}\left\lbrack
         \left(R_s/ r_c\right)^3 -1-3\ {\rm ln} \left( r_i/ r_c\right)
         \right\rbrack &for $w=3/2$,\cr} 
\eeq
where $r_i$ is the location of the front, and $\beta= (3-2w)^{-1}$. This 
defines a critical exponent, corresponding to the maximum density gradient that is able to stop the ionization front, 
\beq
w_{f}= {3\over 2}\left\lbrack 1-\left( r_c\over R_s\right)^3
              \right\rbrack^{-1},
\eeq
and above which no initial HII radius exists. Note that for $R_s/r_c>2$ the 
critical value becomes $w_f\simeq 3/2$. Obviously, the concept of a critical
exponent is not restricted to power-law stratifications and can also be applied
to other types of density distributions, for instance, exponential, gaussian, 
and sech$^2$ profiles (see Franco \etal 1989).

For $w\not= 3/2$, the initial HII region radius can be written as
\beq 
 R_w=g(w)R_s, 
\eeq
with
\beq 
 g(w)= \biggl\lbrack {{{3-2w}\over 3}+{2w\over3}\left({ r_c\over R_s}
        \right)^3}\biggr\rbrack ^{\beta} \left( R_s\over r_c
        \right)^{2w\beta } , 
\eeq
where $R_s$ is the Str\"omgren radius for the density $n_c$.
The solution for $w=3/2$ is
\beq 
 R_{3/2}= r_c\ {\rm exp}\left\{ {1\over 3}\biggl\lbrack\left(
            R_s\over r_c\right)^3-1\biggr\rbrack \right\}.
\eeq

After the formation phase has been completed in clouds with $w\leq w_f$, the 
HII region begins its expansion phase. For simplicity, we assume that the shock
evolution starts at $t=0$ when $R_w$ is achieved. Given that the expansion is 
subsonic with respect to the ionized gas, the density structure inside the HII 
region can be regarded as uniform and its average ion density at time $t$ is
\beq
 \rho_i(t)\simeq \mu_i {(9-6w)^{1/2}\over 3-w} (2n_c)
          R_s^{3/2}R^{-3/2}(t),
\eeq
where $\mu_i$ is the mass per ion, and $R(t)$ is the radius of the HII region 
at the time $t$. For $w\leq 3/2$, the radius can be approximated by
\beq
 R(t)\simeq R_w \left\lbrack 1 + {7-2w\over 4}
         \left( 12\over 9-4w\right)^{1/2} {c_it\over R_w}
         \right\rbrack^{4/(7-2w)}, 
\eeq
where $c_i$ is the sound speed in the ionized gas. The ratio of total mass 
(neutral plus ionized), $M_s(t)$, to ionized mass, $M_i(t)$, contained within 
the expanded radius evolves as
\beq
 {M_s(t)\over M_i(t)}\simeq \left\lbrack R(t)\over R_w
          \right\rbrack^{(3-2w)/ 2}.
\eeq
This equation indicates: i) for $w<3/2$, the interphase between the ionization 
front and the leading shock accumulates neutral gas and its mass grows with 
time to exceed even the mass of ionized gas, and ii) for $w=3/2$= $w_{crit}$, 
the two fronts move together without allowing the formation and growth of a 
neutral interphase. Note that the decreasing ratio predicted by equation (28)
for $w>3/2$ is physically meaningless and it only indicates that the ionization 
front overtakes the shock front (and proceeds to ionize the whole cloud). Thus,
regardless of the value of the critical exponent for the formation phase, 
$w_f$, the expansion phase is characterized by a critical exponent with a well 
defined value, $w_{crit}=3/2$, which is independent of the initial conditions.
Furthermore, this critical exponent $w_{crit}=3/2$ is not affected by dust 
absorption (see Franco \etal 1990).

For $3/2<w<w_f$, the ionization front overtakes the shock and the whole cloud 
becomes ionized. In this case, the pressure gradient simply follows the density
gradient. The ionized cloud is set into motion, but the expanded core (now 
with a radius identical to the position of the overtaken shock) is the densest 
region and feels the strongest outwards acceleration. Then, superimposed on the 
general gas expansion there is a wave driven by the fast growing core (the wave 
location defines the size of the expanded core), and the cloud experiences the 
so-called ``champagne" phase. This core expansion tends to accelerate with time 
and two different regimes, separated by $w=3$, are apparent: a {\it slow} 
regime with almost constant expansion velocities, and a {\it fast} regime with
strongly accelerating shocks.
The slow regime corresponds to $3/2<w< 3$ and the core grows approximately as
\beq
 r(t)\simeq r_c + \left\lbrack 1+\left(3\over 3-w\right)^{1/2}
               \right\rbrack { c_it}, 
\eeq
where for simplicity the initial radius of the denser part of the cloud
has been set equal to $r_c$, the initial size of the core.
For $w=3$ the isothermal growth is approximated by
\beq
 r(t)\simeq 3.2r_c \left\lbrack c_it\over r_c\right\rbrack^
               {1.1}. 
\eeq
For $w>3$, the fast regime, the shock acceleration increases with
increasing values of the exponent and the core expansion is approximated by
\beq
 r(t)\simeq r_c \left\lbrack 1+\left(4\over w-3\right)^{1/2}
       \left(\delta+2-w\over 2\right){c_it\over
       r_c}\right\rbrack^{2/(\delta+2-w)},
\eeq
where $\delta \simeq 0.55(w-3)+2.8$.

\subsection{Wind-driven Bubbles}

The evolution of the cavity created by a stellar wind, a wind-driven bubble,
can also be derived with the thin shell approximation described above. The 
thermalization of the wind creates a hot shocked region enclosed by two 
shocks: a reverse shock that stops the supersonic wind, and an outer shock that
penetrates the ambient gas. The gas processed by each shock is separated by a 
contact surface, the contact discontinuity. The kinetic energy of the wind is 
transformed into thermal energy at the reverse shock producing a hot gas (\eg
Weaver \etal 1977). For the case of a strong wind evolving in a high-density 
and dusty molecular core, the properties of the cooling are poorly known, but
the shocked ambient 
gas cools down very quickly and a thin external shell is formed on time-scales 
of the order of years. Thus, the ambient gas is collected in a thin shell by
the outer shock during most of the evolution. The case of the shocked stellar
wind is less clear because the cooling time there can be substantially longer 
than in the shocked ambient gas, and it is difficult to define when the thin
shell is formed behind the reverse shock. Thus, one can simply derive the
limits for the evolution of the reverse shock in both the adiabatic and 
radiative modes.

The density in a steady wind, with a constant mass loss, decreases as 
\beq
\rho_w = {\dot{M} \over 4 \pi r^2 v_{\infty}}, 
\eeq
where $\dot{M}$ is the stellar mass-loss rate, and $v_{\infty}$ is the wind 
speed. The pressure in the shocked wind region is defined by the wind ram 
pressure, $\rho_w v_{\infty}^2$, at the location of the reverse shock and is
given by
\beq
P_i= {\dot{M} v_{\infty} \over 4 \pi R_{rs}^2 } , \label{rampress}
\eeq
where $R_{rs}$ is the radius of the reverse shock.

\subsubsection{Adiabatic Case}

For an adiabatic bubble evolving in a constant density medium and powered by a
constant mechanical luminosity, $L_{\rm w}$, the thermal energy of the shocked 
wind region grows linearly with time, $E_{th} = 5\,L_{\rm w} t/11$ (Weaver 
\etal 1977). The shocked ambient medium is concentrated in the external thin
shell and the the bubble radius, $R_b$ is the radius of the contact 
discontinuity. The thermal pressure of the bubble interior changes as $P_i = 
(5L_{\rm w} t)/(22\pi R_b^3)$. This pressure is equal to that given by equation 
(\ref{rampress}), and the locations of the reverse and outer shocks are
related by mass conservation
\beq
R_{rs}= R_b^{3/2} \left( {11 \dot{M} v_{\infty} \over 10 L_{\rm w} \,\,t}
\right)^{1/2} .  \label{shocks}
\eeq
The solution for $R_b$, then, also provides the evolution of the reverse shock.
The initial radius in this case is very small, and we simply set $R_b=R_0
(t/t_0)^{\beta}$, where $R_0$ now represents the bubble radius at some 
reference time $t_0$. Again, as in the HII region case, the formal solution is
\beq 
R_b = \int_0^t \left[ \frac{3 \beta}{(4 \beta -1) \rho_0} \left(
\frac{5 L_{\rm w} t^{1 - 3 \beta} t_0^{3\beta}}{22 \pi R_0^3} - P_0 \right) \right]^{1/2}
\,\,dt .
\eeq

At early times, $P_i >> P_0$ and $P_0$ can be neglected in the above equation. 
The exponent is then defined by $\beta = 1+ (1-3\beta )/2$, giving the well 
known adiabatic expansion in a medium with constant density $R_b\propto L_{\rm 
w}^{1/5}\rho_0^{-1/5} t^{3/5}$. The position of the reverse shock (equation 
\ref{shocks}), then, evolves as $R_{rs}\propto t^{2/5}$. With this time
dependence, the internal pressure drops as $P_i \propto t^{-4/5}$. At later 
times, when $P_i\rightarrow P_0$, the bubble reaches quasi-equilibrium with
the ambient gas
and the growth is $R_{b,{\rm eq}}\propto (L_{\rm w}/P_0)^{1/3} t^{1/3}$.
In pressure equilibrium, the radius of an $adiabatic$ bubble grows at a 
slow rate, but {\it no steady state} solution exists during this 
stage. The final radius of the reverse shock is simply given by the 
balance between the external and the wind ram pressures. The wind 
density at equilibrium is $\rho_{w,{\rm eq}} = P_0 / v_{\infty}^2$, and
the location of the reverse shock is
\beq 
R_{rs,{\rm eq}}=\left[ \frac{\dot{M} \,v_{\infty}}{4 \,\pi \,P_0} \right]^{1/2}
\simeq 2.3\times 10^{-2} \left[ \frac{\dot{M_6} \,\,v_{\infty,8}}{P_7}
\right]^{1/2}  \ \  \,\,{\rm pc} , \label{Req} 
\eeq
where $\dot{M_6}=\dot{M}/10^{-6}$ M$_\odot$ yr$^{-1}$, and $v_{\infty,8}= 
v_{\infty}/10^8$ cm s$^{-1}$. Using mass conservation in the shocked 
wind region, $M_{sw}=\dot{M}t$, the quasi-equilibrium radius of an {\it 
ultracompact} wind-driven bubble is given by
\beq
R_{b,{\rm eq}}= R_{rs,{\rm eq}} \left( 1+ \frac{3 L_w t}
{8 \pi P_0 R_{rs,{\rm eq}}^3} \right)^{1/3} .
\eeq 
Clearly, for adiabatic bubbles, $\beta$ goes from $3/5$ at early times to $1/3$ 
at late times. 

\subsubsection{Radiative Case}
 
Once radiative losses become important, the hot gas looses its pressure and 
the bubble collapses into a simple structure: the free-expanding wind collides 
with the cold shell and the gas is thermalized and cools down to low 
temperatures in a cooling length. At this moment, the shell
becomes static at the radius $R_{rs,{\rm eq}}$.

If the bubble becomes radiative before the final radius, $R_{rs,{\rm eq}}$, is
reached, the shell is pushed directly by the wind pressure. For this radiative
bubble case, the formal solution is
\beq 
R_b = \int_0^t \left[ \frac{3 \beta}{(4 \beta -1) \rho_0} \left(
\frac{\dot{M} \,v_{\infty} t_0^{2\beta}}{4 \pi R_0^2 t^{2\beta}}
 - P_0 \right) \right]^{1/2}
\,\,dt .
\eeq
The early times solution, with $P_i>>P_0$, gives the relation $\beta =1-\beta$,
and one recovers the well known solution for radiative bubbles $R_b\propto 
t^{1/2}$ (\eg Steigman \etal 1975). The ram pressure now evolves as $P_i 
\propto t^{-1}$. At late times, one gets $\beta \rightarrow 0$ and the shell 
reaches the final equilibrium radius (eqn. \ref{Req}). Thus, the exponent in 
this case evolves from 1/2 to 0. Note that if the bubble starts on an 
adiabatic track and becomes radiative before pressure equilibrium is achieved, 
the exponent varies from 3/5 to 1/2, and then to zero. Figure 3 illustrates
the evolution of an ultracompact wind-driven shell in a high-pressure core.
\begin{figure}
\vspace*{66mm}
\begin{minipage}{66mm}
\includegraphics{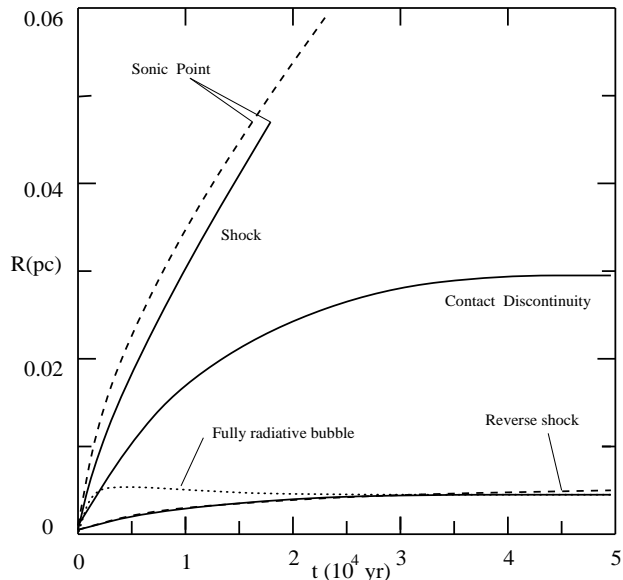}
\end{minipage}

\caption{The evolution of an ultracompact wind-driven bubble, with $\dot{M_6}=
1$ and $v_{\infty,8}= 1$, in a high-pressure core, with $P_7=1$.}
\label{fig:3}
\end{figure}

\section{Discussion}

\subsection{Young Stars}

The history of the pressure in a star forming region, then, follows a
somehow simple scheme. The initial properties and pressure of the gas in a star
forming cloud is defined by self-gravity. Once young stars appear, their 
energy input modifies the structure and evolution of the cloud. 
This is particularly true for massive stars, 
their radiative and mechanical energy inputs 
are even able to 
disrupt their parental clouds.
In the case of the dense cloud cores, the sizes of either HII
regions or wind-driven bubbles are severely reduced by the large
ambient pressure (Garc\'{\i}a-Segura \& Franco 1996). In fact, the
pressure equilibrium radii of ultra-compact HII regions are actually
indistinguishable from those of ultra-compact wind-driven bubbles, and 
they could be stable and long lived. Actually, Xie \etal\ (1996) have 
recently found 
evidence indicating that the smaller UCHII seem
to be embedded in the higher pressure cores.

The situation is completely different when the stars are located near the edge of the cloud core. The resulting HII regions (and also wind-driven bubbles) 
generate supersonic outflows. Cases with $w>w_{crit}$ lead to the champagne phase: once the
cloud is fully ionized, the expansion becomes supersonic. 
For spherical clouds with a small constant-density core
and a power-law density distribution, $r^{-w}$, outside the core,
there is a critical exponent ($w_{crit}=3/2$) above which the cloud becomes
completely ionized. 
This represents an efficient mechanism for cloud
destruction and, once the parental molecular cloud is completely ionized, can
limit the number of massive stars and the star formation rate (Franco \etal
1994).  For a cloud of mass $M_{GMC}$, with only 10\% of this mass 
concentrated in star-forming dense cores, the number of newly formed OB
stars required for complete cloud destruction is
\begin{equation}
N_{OB} \sim 30 
\frac{M_{GMC,5}n_{3}^{1/5}}{F_{48}^{3/5}(c_{i,15} t_{MS,7})^{6/5}}.
\end{equation}
where $M_{GMC,5}=M_c/10^5$ \Mo, $n_{3}=n_{0}/10^3$ cm$^{-3}$,
$c_{i,15}=c_{i}/15$ km s$^{-1}$, and $t_{MS,7}$ is the main sequence
lifetime in units of $10^7$ yr.
This corresponds to a total star forming
efficiency of about $\sim 5$ \% (larger average densities and cloud masses can
 result in higher star formation efficiencies).

Summarizing, photoionization from OB stars can destroy the parental
cloud in relatively short time scales, and defines the limiting number
of newly formed stars. The fastest and most effective destruction
mechanism is due to peripheral, blister, HII regions, and they can
limit the star forming efficiency at galactic scales. Internal HII
regions at high cloud pressures, on the other hand, result in large
star forming efficiencies and they may be the main limiting mechanism
in star forming bursts and at early galactic evolutionary stages (see
Cox 1983).

\subsection{Slow Winds from Evolved Stars}

The evolution of evolved stellar winds in high-pressure regions is discussed 
in Franco \etal (1996), and here we just repeat the relevant parts.
As the cloud is dispersed, the average gas density decreases and the
newly formed cluster becomes visible. The individual HII regions merge
into a single photo-ionized structure and the whole cluster now powers
an extended, low density, HII region. The stellar wind bubbles now can
grow to larger sizes and some of them begin to interact. As more winds
collide, the region gets pressurized by interacting winds and the
general structure of the gas in the cluster is now defined by this
mass and energy input (Franco \etal\ 1996).

Given a total number of massive stars in the cluster, $N_{OB}$, and their 
average mass input rate, $<\dot{M}>$, the pressure due to interacting 
adiabatic winds is
\begin{equation}
P_i\sim \frac{N_{OB} <\dot{M}> c_i}{4 \pi r^2_{clus}}\sim 10^{-8}
\frac{N_{2} <\dot{M}_6> c_{2000}}{r^2_{pc}} \ \ \ \ {\rm dyn \ cm^{-2}},
\end{equation}
where $r_{pc}=r_{clus}/1$ pc is the stellar group radius, $N_{2}=N_{OB}/10^2$, 
$<\dot{M}_6>=<\dot{M}>/10^{-6}$ M$_\odot$ yr$^{-1}$, and
$c_{2000}=c_i/2000$ km s$^{-1}$ is the sound speed in the interacting
wind region. This is the central pressure driving the expansion of the
resulting superbubble before the supernova explosion stage. For modest
stellar groups with relatively extended sizes, like most OB
associations in our Galaxy, the resulting pressure is only slightly
above the ISM pressure (\ie for $N_{2} \sim 0.5$ and $r_{pc}\sim 20$,
the value is $P_i \sim 10^{-11}$ dyn cm$^{-2}$).  For the case of rich
and compact groups, as those generated in a starburst, the pressures
can reach very large values. For instance, for the approximate cluster
properties in starbursts described by Ho (1996), $r_{pc}\sim
3$ and $N_{2}>10$, the resulting pressures can reach values of the
order of $P_1\sim 10^{-7}$ dyn cm$^{-2}$, similar to those due to
self-gravity in star forming cores. At these high pressures, the wind of a
red giant (or supergiant) cannot expand much and the bubble reaches 
pressure equilibrium at a relatively small distance from the evolving star. 
Thus, the large mass lost during the slow red giant 
wind phase is concentrated in a dense circumstellar shell.
\begin{figure}
\vspace*{43mm}
\begin{minipage}{43mm}
\includegraphics{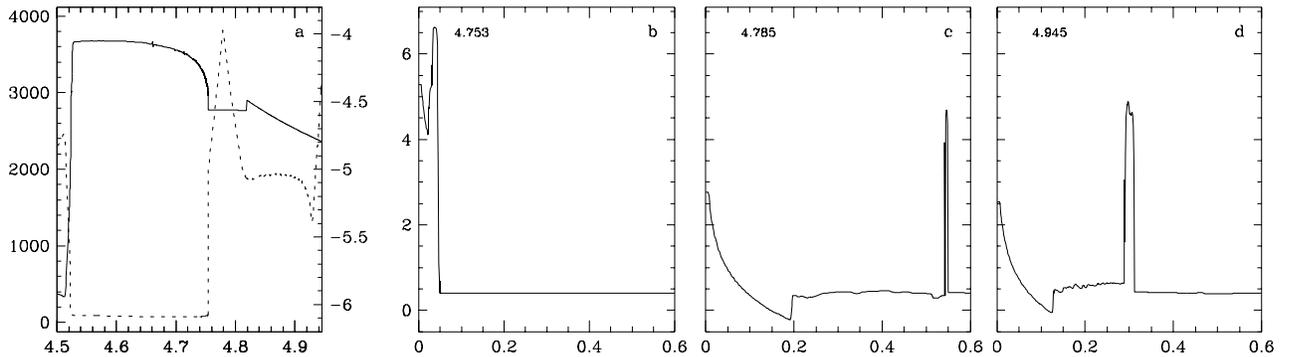}
\end{minipage}

\caption{
Wind-driven bubble from an evolved 35\,M$_{\odot}$ star in a high pressure 
medium. {\bf a)} Left scale: terminal wind velocity (km s$^{-1}$); right scale:
mass loss rate (M$_{\odot}$ yr$^{-1}$); horizontal scale: time
(millions of yr). {\bf (b)-(d)}: Evolution of the wind-driven bubble. The gas
density (cm$^{-3}$) is plotted in a logarithmic scale versus the radial 
distance (pc). Evolutionary times (upper-left corner of each 
panel) are given in million years.}
\label{fig:4}
\end{figure}
Figure 4 shows the evolution of a wind-driven
bubble around a 35\,M$_{\odot}$ star. Fig. 4a shows the wind 
velocity and mass-loss rate (dashed and solid lines, respectively:
Garc\'{\i}a-Segura, Langer \& Mac Low 1996). The simulations are done only over the 
time spanning the red supergiant and Wolf-Rayet phases, and assume that the
region is already pressurized by the main sequence winds from massive stars. 

We have used the AMRA code, as described by Plewa \& R\'o\.zyczka (1996).  
During the RSG phase the wind-driven shell is located very close ($R\approx 0.04$
pc) to the star due to a very low wind ram-pressure from the wind
(Fig. 4b). Later on 
(Fig. 4c), the powerful WR wind pushes the shell away from the star to the 
maximum distance of $R\approx 0.54$ pc. Still later, when the wind has 
variations, the shell adjusts its position accordingly, and reaches the 
distance $R\approx 0.3$ pc at the end of simulation (Fig. 4d). It must be stressed 
that the series of successive accelerations and decelerations of the shell 
motion during the WR phase will certainly drive flow instabilities and cause 
deviations from the sphericity assumed in our model. The role of these 
multidimensional instabilities in the evolution of the shell is currently 
under study (with 2-D and 3-D models), and the results will be presented in 
a future communication.
Regardless of the possible shell fragmentation, however, when the star 
explodes as a supernova, the ejecta collides with a dense circumstellar 
shell. This interaction generates a bright and compact supernova remnant, with 
a powerful photoionizing emission (\ie Terlevich \etal\ 1992; Franco \etal\ 
1993; Plewa \& R\'o\.zyczka 1996), that may also be a very strong 
radio source, like SN 1993J (see Marcaide \etal 1995). If the shell is 
fragmented, the ejecta-fragment interactions will occur during a series of 
different time intervals, leading to a natural variability in the emission at 
almost any wavelength (see Cid-Fernandes \etal 1996). This type of interaction 
is also currently under investigation, and further modeling will shed more 
light on the evolution of SN remnants in high-pressure environs.

JF and GGS acknowledge partial support from DGAPA-UNAM grant IN105894, 
CONACyT grants 400354-5-4843E and 400354-5-0639PE, and a R\&D Cray research 
grant. The work of TP was partially supported by the grant KBN 2P-304-017-07 
from the Polish Committee for Scientific Research. The simulations were 
performed on the CRAY Y-MP of the Supercomputing Center at UNAM, and 
on a workstation cluster at the Max-Planck-Institut f\"ur Astrophysik.

\end{document}